\documentclass[]{article}
\usepackage{latexsym}
\usepackage{amsmath}
\usepackage{amssymb}
\usepackage{amsfonts}
\usepackage[mathcal]{euscript}
\usepackage{graphicx}
\usepackage{epsfig}
\usepackage{array}

\newtheorem{theorem}{Theorem}
\newtheorem{lemma}{Lemma}

\newtheorem{ExampleDef}{Example}[section]

\newcommand{\Example}[3]{
  \begin{list}{}{
      \setlength{\leftmargin}{1em}}     
    \item                               
    \small                              
    \begin{ExampleDef} \rm              
      {\bf \hspace{-1ex}: #1}           
      #2                                
      \hfill {\large \boldmath $\Box$}  
      \label{ex:#3}                      
    \end{ExampleDef}
  \end{list}}

\begin{document}
\begin{center}
{\Large {\bf Where To Go and How To Go: 
     a Theoretical Study of \\ Different Leader Roles in Networked Systems\par}}
\vspace{1.0em}
{\large Wei Wang and Jean-Jacques E. Slotine \par} 
{Nonlinear Systems Laboratory \\
Massachusetts Institute of Technology \\
Cambridge, Massachusetts, 02139, USA 
\\ wangwei@mit.edu, \ jjs@mit.edu 
\par}
\vspace{2em}
\end{center}

\begin{abstract}
This letter studies synchronization conditions for distributed dynamic
networks with different types of leaders. The role of a leader
specifying a desired global state trajectory through local interactions
(the power leader) has long been recognized and modeled. This paper
introduces the complementary notion of a 'knowledge' leader holding
information on the target dynamics, which is propagated to the entire
network through local adaptation mechanisms.  Knowledge-based
leader-followers networks have many analogs in biology, e.g. in
evolutionary processes and disease propagation. Different types of
leaders can co-exist in the same network.
\end{abstract}

%
\section{Introduction}
Recent results in the theoretical study of synchronization and group
cooperation~\cite{kumar, jadbabaie03, leonard01, olfati_1, pikovsky,
wei03-2, strogatz03, vicsek02, wei03-1} have greatly helped in
understanding distributed networked systems in the nature world. In
these systems, each element can only get local information from a set of
neighbors and the whole network is able to present a collective
behavior. Examples of such networked systems pervade nature at every
scale, including neural networks,
pacemaker cells, flashing fireflies, chirping crickets, and the
aggregate motions of bird flocks, fish schools, animal herds and bee
swarms, just to cite a few.  
For a diffusion-coupled network with arbitrary size and
general structure, \cite{wei03-2, wei03-1} provides an explicit
synchronization condition by setting a coupling-strength threshold,
which is computed based on network connectivity and uncoupled element
dynamics.

For a leaderless network composed of peers, the phase of its
collective behavior is hard to predict, since it depends on the
initial conditions of all the coupled elements. Thus, for the whole
network to behave as desired, an additional group leader is necessary. 
Here the leader is defined as the one whose
dynamics is independent and thus followed by all the others. Such a
leader-followers network is especially popular in natural aggregate
motions, where the leader ``tells'' the followers ``where to go''.  We
name this kind of leader the {\it power leader}. The synchronization
condition for a dynamic network with a power leader was derived
in~\cite{wei03-2, wei03-1} and will be briefly reviewed here.

In this letter, we introduce a new leader role, which we call a {\it
knowledge leader}. In a knowledge-based network, members' dynamics are
initially non-identical and mutually coupled. 
The leader is the one whose dynamics is fixed
or changes comparatively slowly. The followers obtain dynamics
knowledge from the leader through adaptation. In this sense, a
knowledge leader can be understood as the one who indicates ``how to
go''.  In fact, knowledge leaders may exist in many natural processes.
For instance, in evolutionary biology~\cite{nowak03, nowak02},
the adaptive model we describe could represent genotype-phenotype
mapping. Similar is infectious-disease dynamics~\cite{may}. Knowledge leaders may
also exist in animal aggregate motion as a junior or
injured member with limited capacities.  Different than a power leader,
a knowledge leader does not have to be dynamically independent. It may
be located at any position in a network. Using Lyapunov analysis, we will
derive the conditions of synchronization and also dynamics-convergence 
for knowledge-based networks. We will then show that different types of
leaders can co-exist in the same network.

%
\section{Power Leader} \label{sec:power-leader}
\begin{figure}[h]
\begin{center}
\epsfig{figure=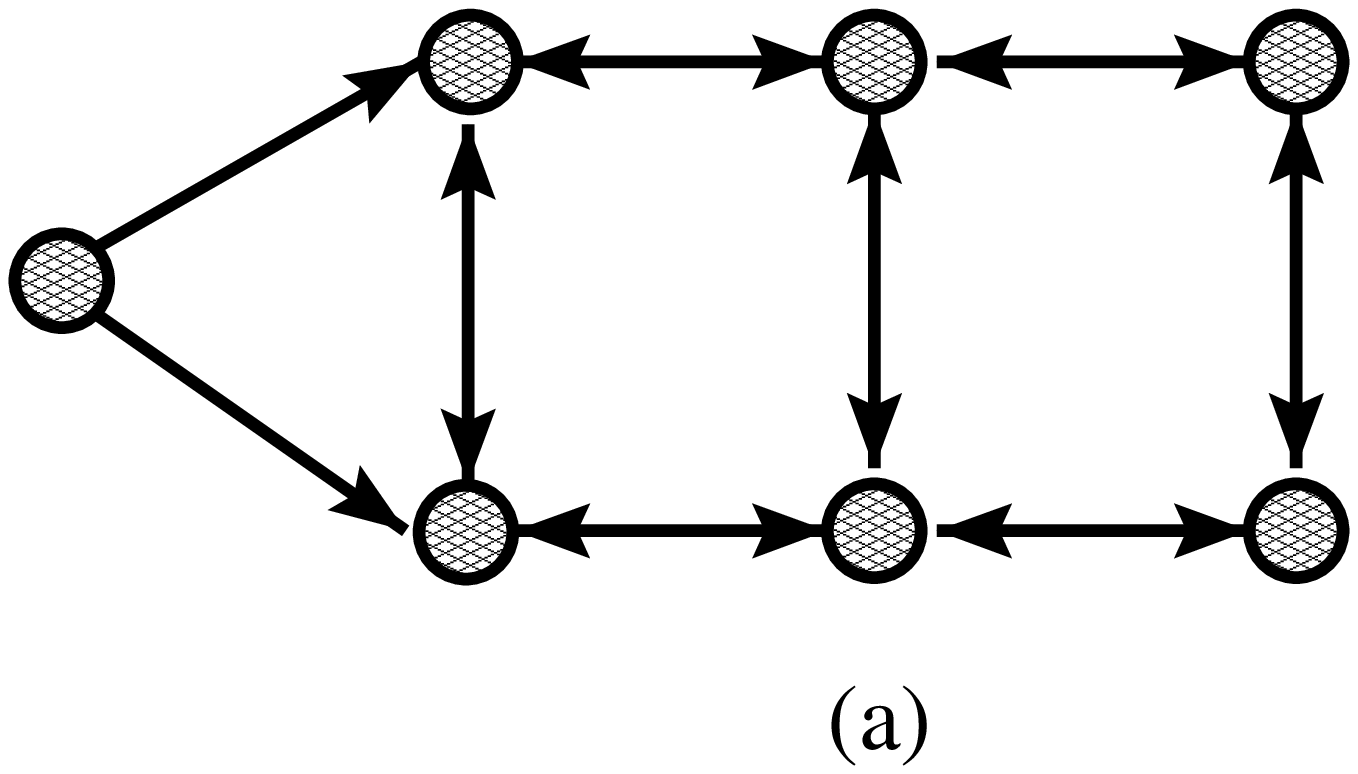,height=25mm,width=40mm}
\ \ \ \ \ \ \ \ 
\epsfig{figure=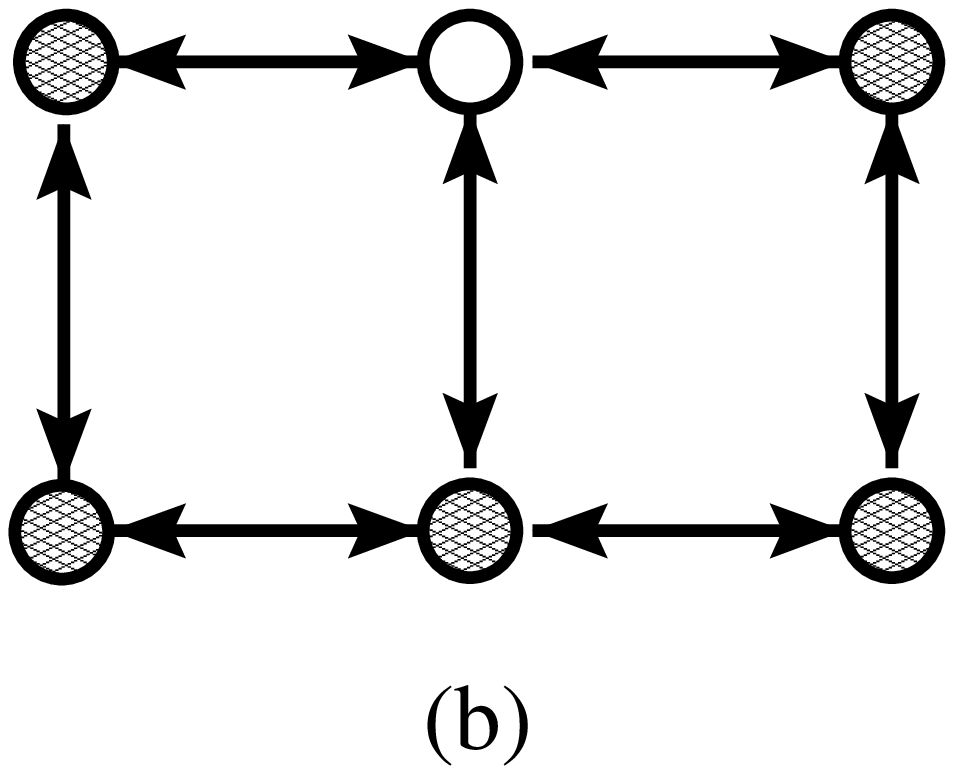,height=25mm,width=30mm}
\ \ \ \ \ \ \ \ 
\epsfig{figure=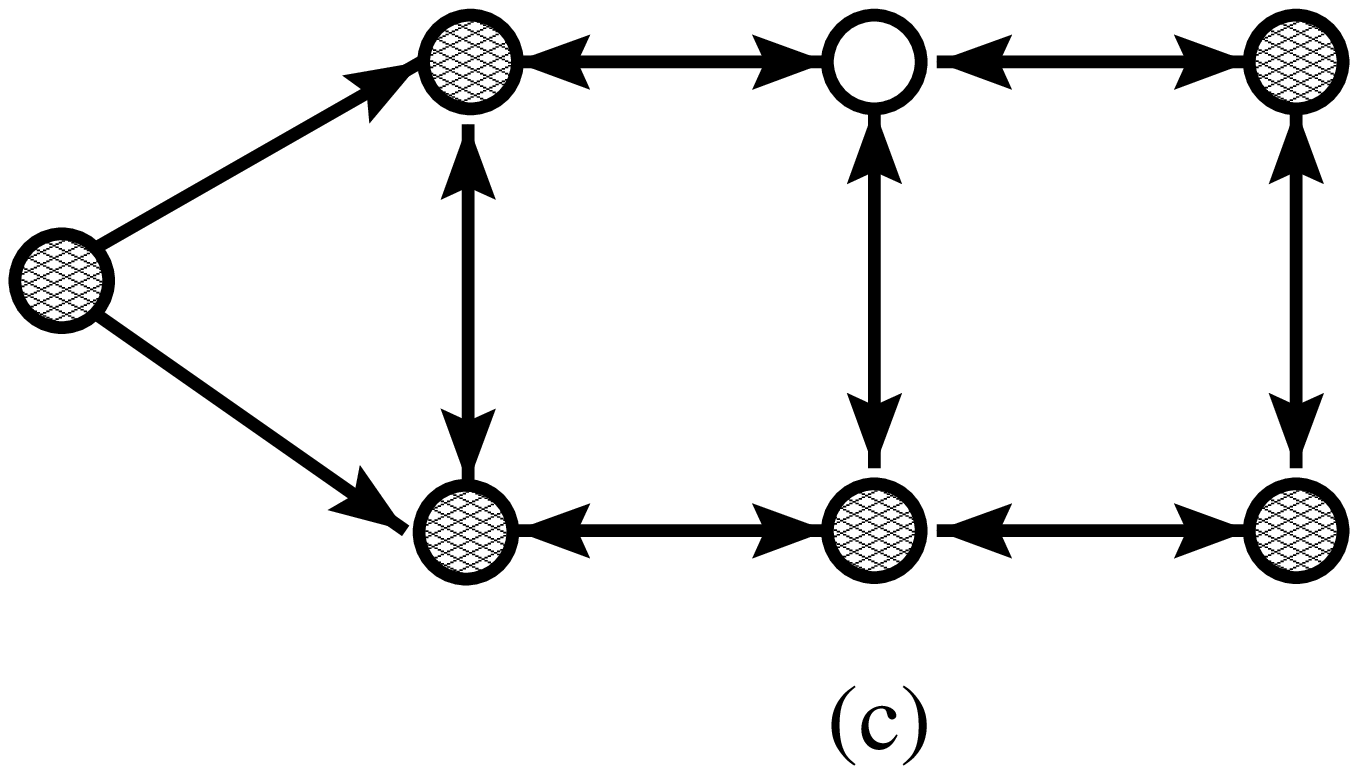,height=25mm,width=40mm}
\caption{The graphs illustrate networked systems with
(a). a power leader (the most left node);
(b). a knowledge leader (the hollow node);
(c). both leaders. The arrows indicate the
directions of the couplings.}
\label{fig:power-leader}
\end{center}
\end{figure}

Consider the dynamics of a coupled network containing one power leader 
and $n$ power followers 
\begin{eqnarray}
\dot{\bf x}_0 &=&  {\bf f}({\bf x}_0,t)  \label{eq:leader-followers} \\
\dot{\bf x}_i &=&  {\bf f}({\bf x}_i,t) + 
\sum_{j \in {\mathcal N}_i} {\bf K}_{ji}\ ({\bf x}_j - {\bf x}_i) 
                 + \gamma_i\ {\bf K}_{0i}\ ({\bf x}_0 - {\bf x}_i) 
\ \ \ \ i=1,\ldots,n  \nonumber
\end{eqnarray}
Here vector ${\bf x}_0  \in \mathbb{R}^m$ is the state of the 
leader whose dynamics is
independent, and ${\bf x}_i$ the state of the $i$th follower. 
Vector function ${\bf f}$ represents the uncoupled dynamics,
which is assumed to be identical for each element. 
For notation simplicity, the coupling forces are set to be diffusive,
where all coupling gains are symmetric positive definite,
and the couplings between the followers are bi-directional with
$
{\bf K}_{ji} = {\bf K}_{ij}
$
if both $i, j \ne 0$.
${\mathcal N}_i$ denotes the set of peer-neighbors of element $i$,
which for instance could be defined as the set of the followers
within a certain distance around element $i$. 
$\gamma_i$ is equal to either $0$ or $1$, representing the connection
from the leader to the followers.
In our model, the network connectivity can be very general. 
Thus ${\mathcal N}_i$ and $\gamma_i$ can be defined arbitrarily.
An example is illustrated in Figure~\ref{fig:power-leader}(a).

Since the dynamics of ${\bf x}_0$ is independent, it can be
considered as an external input to the rest of the network.
Therefore it can be seen as if it  
contains only the followers and is an undirected graph with
$n$ nodes. We further assume that it has $\tau$ inner links.
\begin{theorem}\label{th:power-leader} 
The states of all the followers will converge exponentially to the state of
the leader if
$$
 \displaystyle
\lambda_{min} ( {\bf L}_{\bf K} + {\bf I}^n_{\gamma_i {\bf K}_{0i}} )
 > \max_{i=1}^n 
\lambda_{max}(\frac{\partial {\bf f}}{\partial {\bf x}}({\bf x}_i,t) )_s
\ \ \ \mathrm{uniformly.}
$$
\end{theorem}
In Theorem~\ref{th:power-leader}, $\lambda$ represents an eigenvalue and
subscript $s$ the symmetric part of a matrix;
notation ${\bf I}^n_{\gamma_i {\bf K}_{0i}}$ denotes an $n \times n$ 
block diagonal matrix with the $i^{\mathrm{th}}$ diagonal entry as
$\gamma_i {\bf K}_{0i}$; ${\bf L}_{\bf K}$ is the weighted Laplacian 
matrix~\cite{godsil} and
$$
{\bf L}_{\bf K}\ =\ {\bf D}\ {\bf I}^{\tau}_{{\bf K}_{ij}} \ {\bf D}^T
$$ 
where the $n \times \tau$ block matrix ${\bf D}$ is a
generalized incidence matrix by replacing each number $1$ or $-1$
in the incidence matrix~\cite{godsil} with identity matrix 
${\bf I} \in \mathbb{R}^{m \times m}$ or $-{\bf I}$. Note that the
incidence matrix is defined by
assigning an arbitrary orientation to the undirected graph.
${\bf I}^{\tau}_{{\bf K}_{ij}}$ is
a $\tau \times \tau$ block diagonal matrix with the $k^{\rm{th}}$ diagonal entry 
${\bf K}_{ij}$ corresponding to the weight of the $k^{\rm{th}}$ link
which connects the nodes $i$ and $j$. The proof of Theorem~\ref{th:power-leader}
is based on Contraction Theory~\cite{winni98, winnithesis}, 
the details of which can be found in~\cite{wei03-2, wei03-1}.

\noindent {\it A few remarks}: 
\newline
\indent $\bullet$\ Theorem~\ref{th:power-leader} can be 
extended to study networks
with unidirectional couplings between the followers, with
positive semi-definite couplings, or with switching 
structures~\cite{wei03-2, wei03-1}. Moreover, 
as a generalized understanding, the network does not have to have only one leader.
It can be a group of leading elements. The leader even does not have to be independent. It
can receive feedback from the followers as well. Such an example is 
synchronization propagation~\cite{wei03-2}, where the density is not smoothly
distributed through the whole network. Since synchronization rate 
depends on both coupling strengths and network connectivity, 
a high-density region will synchronize very quickly despite disturbances from 
other parts of the network. The inputs from 
these leaders then facilitate synchronization in low-density regions.
\newline
\indent $\bullet$\ Note that different leaders $ \ {\bf x}_0^j \ $ of arbitrary
dynamics can define different {\it primitives} which can be
combined. Contraction of the follower dynamics ($i = 1,\ldots,n$)
$$
\dot{\bf x}_i =  {\bf f}({\bf x}_i,t)\ + \ 
\sum_{j \in {\mathcal N}_i} {\bf K}_{ji}\ ({\bf x}_j - {\bf x}_i) 
\ + \ 
\sum_j \alpha_j (t) \ \gamma_i^j \ {\bf K}_{0i}^j\ ({\bf x}_0^j - {\bf x}_i)
$$
is preserved if $\ \sum_j \alpha_j(t) \ge 1, \ \forall t \ge 0$.
\newline
\indent $\bullet$\ Besides its dubious moral implications, 
Theorem~\ref{th:power-leader}
also means that it is easy to detract a group from its nominal behavior
by introducing a ``covert'' element, with possible applications to
group control games, ethology, and animal and plant mimicry.
\newline
\indent $\bullet$\ Besides orientation, the moving formation with a power
leader has other advantages, such as energy saving in
aerodynamics~\cite{cutts, seiler}.

%
\section{Knowledge Leader} \label{sec:knowledge-leader}
A knowledge-based leader-followers network is composed of elements with initially 
non-identical dynamics. A knowledge leader may be located in any position inside a
network as we illustrated in Figure~\ref{fig:power-leader}(b).
Its dynamics is fixed or slowly changing, while those of the followers 
are learned from the leader through adaptation. If we consider
the power leader as the one which tells the rest of the network ``where to go'', a 
knowledge leader indicates ``how to go''. Synchronization or group agreement
can still be achieved in such a network with only local interactions. 

Consider a coupled network containing $n$ elements without a power leader
\begin{equation} \label{eq:general-network-knowledge-leader}
\dot{\bf x}_i =  {\bf f}({\bf x}_i,{\bf a}_i, t) + 
\sum_{j \in {\mathcal N}_i} {\bf K}_{ji} ( {\bf x}_j - {\bf x}_i)
\ \ \ \ i=1,\ldots,n
\end{equation}
where the connectivity can be general.
Assume now that the uncoupled dynamics ${\bf f}({\bf x}_i,{\bf a}_i, t)$ 
contains a parameter 
set ${\bf a}_i$ which has a fixed value ${\bf a}$ for all the knowledge leaders. 
Denote $\Omega$ as the set of the followers, whose adaptation laws are based
on local interactions
\begin{equation} \label{eq:adaptive-law-network}
\dot{\bf a}_i\ =\ {\bf P}_i {\bf W}^T({\bf x}_i,t) 
 \sum_{j \in {\mathcal N}_i} {\bf K}_{ji}\ ({\bf x}_j - {\bf x}_i)  
\ \ \ \ \forall\ i \in \Omega 
\end{equation}
where ${\bf P}_i>0$ is constant and symmetric, and ${\bf W}({\bf x}_i,t)$ is defined as
$$
{\bf f}({\bf x}_i,{\bf a}_i,t)\ =\ {\bf f}({\bf x}_i,{\bf a},t)\ +\ 
{\bf W}({\bf x}_i,t)\tilde{\bf a}_i
$$
with estimation error $\ \tilde{\bf a}_i = {\bf a}_i-{\bf a}\ $. 

To prove convergence, we define a Lyapunov function
$$
V = \frac{1}{2}\ (\ {\bf x}^T {\bf L}_{\bf K} {\bf x}\ +\    
     \sum_{i \in \Omega} \tilde{\bf a}_i^T {\bf P}^{-1}_i \tilde{\bf a}_i\ )
$$ 
where ${\bf x}^T = [{\bf x}_1^T, {\bf x}_2^T, \ldots, {\bf x}_n^T]$, so that
\begin{eqnarray*}
\dot{V} 
&=& {\bf x}^T {\bf L}_{\bf K} \dot{\bf x}\ +\ 
     \sum_{i \in \Omega} \tilde{\bf a}^T {\bf P}^{-1}_i \dot{\bf a}_i \\
&=&   {\bf x}^T {\bf L}_{\bf K}\ (\  
        \left[ \begin{array}{c} {\bf f}({\bf x}_1,{\bf a},t) \\ \ldots \\  
        {\bf f}({\bf x}_n,{\bf a},t) \end{array} \right]
        - {\bf L}_{\bf K}{\bf x}\ ) \\
&=& {\bf x}^T\ (\ {\bf L}_{\bf K \Lambda} - {\bf L}_{\bf K}^2\ )\ {\bf x}
\end{eqnarray*}
where matrix ${\bf L}_{\bf K \Lambda}$ is symmetric and
\begin{equation} \label{eq:l-k-lambda}
{\bf L}_{\bf K \Lambda}\ =\ {\bf D}\ 
({\bf I}^{\tau}_{{\bf K}_{ij}} {\bf I}^{\tau}_{{\Lambda}_{ij}} )_s\ 
                                     {\bf D}^T\ =\
{\bf D}\ {\bf I}^{\tau}_{({\bf K} \Lambda)_{ijs}}\ {\bf D}^T
\end{equation}
Here ${\bf I}^{\tau}_{\Lambda_{ij}} $ is a $\tau \times \tau$ 
block diagonal matrix with the $k^{\rm{th}}$ diagonal entry 
$$
\Lambda_{ij} = \int_0^1 \frac{\partial {\bf f}}{\partial {\bf x}}
            ({\bf x}_j + \chi ({\bf x}_i - {\bf x}_j), {\bf a}, t)\ d \chi
$$ 
corresponding to the $k^{\rm{th}}$ link which has been assigned an orientation 
by the incidence matrix ${\bf D}$. 
${\bf I}^{\tau}_{({\bf K} \Lambda)_{ijs}}$ is defined in a similar manner
with $({\bf K} \Lambda)_{ijs}$ the symmetric part of ${\bf K}_{ij} \Lambda_{ij}$.

To complete the proof, we use the following lemma, which is derived in 
Appendix~\ref{ap:negative-semi-definite-proof}.
\begin{lemma} \label{lm:negative-semi-definite} 
Giving any ${\bf x}^T=[{\bf x}_1^T, {\bf x}_2^T, \ldots, {\bf x}_n^T]$, if
\begin{equation} \label{eq:negative-semi-definite}
\frac{\lambda_{m+1}^2 ( {\bf L}_{\bf K} )}{\lambda_n({\bf L})} > 
\max_k \lambda_{max}({\bf K \Lambda})_{ijs}
\end{equation}
$
{\bf x}^T\ (\ {\bf L}_{\bf K \Lambda}-{\bf L}_{\bf K}^2\ )\ {\bf x} \le 0
$ and the equality is true if and only if 
${\bf x}_1 = {\bf x}_2 = \cdots = {\bf x}_n$.
\end{lemma}
Note that for condition~(\ref{eq:negative-semi-definite}) to be true,
we need a connected network, an upper bounded $\lambda_{max}({\bf K
\Lambda})_{ijs}$, and strong enough coupling strengths. For an
example, if $m=1$ and all the coupling gains are identical with value
$\kappa$, condition~(\ref{eq:negative-semi-definite}) turns to be
$$
\kappa\ >\ \frac{\lambda_n({\bf L})}{\lambda_2^2({\bf L})}\
\max \frac{\partial {\bf f}}{\partial {\bf x}} 
                       ({\bf x}_i, {\bf a}, t)
$$
\begin{theorem}\label{th:adaptation-network}
For a knowledge-based leader-followers network, the states of all the elements 
will converge together asymptotically if condition~(\ref{eq:negative-semi-definite}) 
is verified and all the states are bounded. Furthermore, 
$\forall\ i \in \Omega$, ${\bf a}_i$ will converge to ${\bf a}$ if
\begin{equation}  \label{eq:parameter-converge-condition}
\exists\ \alpha>0, T>0, \ \ \forall t \ge 0\ \ 
\int_t^{t+T} {\bf W}^T({\bf x}_i, r) {\bf W}({\bf x}_i, r) dr \ \ge\ \alpha {\bf I}
\end{equation}
\end{theorem}
\noindent {\bf Proof:}
Condition~(\ref{eq:negative-semi-definite}) means $V$ is non-increasing.
Assuming all the functions are smoothly differentiable, the boundedness of $\ddot{V}$ 
can be concluded if all the states are bounded. 
According to Barbalat's lemma~\cite{jjsbook}, $\dot{V}$ will then tend to $0$ asymptotically,
implying that all the states ${\bf x}_i$ converge together. Hence, 
${\bf W}({\bf x}_i,t)\tilde{\bf a}_i$ will
tend to zero, which leads to the convergence of the followers' parameters under 
condition~(\ref{eq:parameter-converge-condition}). \hfill {\large \boldmath $\Box$}

Theorem~\ref{th:adaptation-network} implies that new elements can be added 
into the network without prior knowledge of the individual dynamics, and that
elements in an existing network have the ability to recover dynamic 
information if temporarily lost. Similar knowledge-based leader-followers 
mechanism may exist in many natural processes. In evolutionary biology, 
knowledge leaders are essential to keep
the evolution processes uninvasible or evolutionary 
stable~\cite{nowak03, nowak02}. In reproduction, for instance, the leaders could be
senior members.  The knowledge-based mechanism may also describe
evolutionary mutation or disease infection~\cite{may}, where the
leaders are mutants or invaders.  Knowledge-based leader-following may
also occur in animal aggregate motions or human social activities. In
a bird flock, for instance, the knowledge leader can be a junior or
injured member whose moving capacity is limited, and which is
protected by others through dynamic adaptation.

Note that the adaptive model we described represents a genotype-phenotype mapping,
where adaptation occurring in genotypic space is based on the interactions
of behavioral phenotypes. Due to its complexity, genotype-phenotype mapping 
remains a big challenge today in evolutionary biology~\cite{nowak03}.

\begin{figure}[h]
\begin{center}
\epsfig{figure=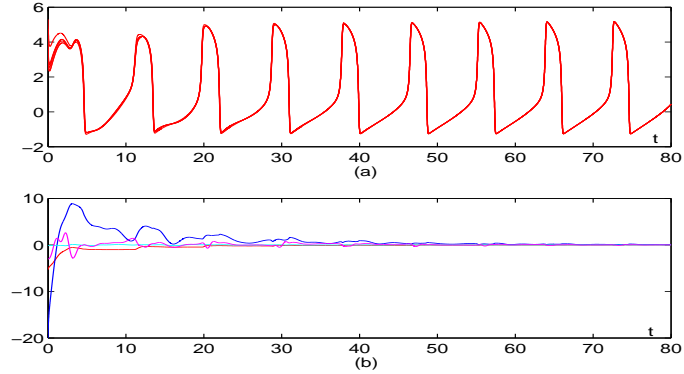,height=50mm,width=90mm}
\end{center}
\caption{Simulation results of Example~\ref{ex:adaptation-fn-network}.  
With initial conditions chosen arbitrarily, the plots show that
(a).states $v_i$ ($i=1,\ldots,6$) synchronize in the time space;
(b).estimation error set $\tilde{\bf a}_i$ of any of the knowledge 
followers vanish in the time space.} 
\label{fig:fn-adp-network}
\end{figure}

\Example{}{Consider six FitzHugh-Nagumo neurons~\cite{fitzhugh, murray, nagumo}, a
famous spiking neuron model, connected as in Figure~\ref{fig:power-leader}(b)
\begin{equation*} 
\begin{cases}  
\displaystyle 
\ \dot{v}_i = v_i(\alpha_i - v_i)(v_i-1)-w_i+I_i + 
             \sum_{j \in {\mathcal N}_i} k_{ij}(v_j-v_i)  \\  
\ \dot{w}_i = \beta_i v_i - \gamma_i w_i 
\ \ \ \ \ \ \ \ \ i=1,\ldots,6
\end{cases}
\end{equation*}
Assume that the parameter set  
$\ {\bf a}_i=[\alpha_i, I_i, \gamma_i, \beta_i]^T\ $
is fixed to the only knowledge leader, and those of the others
change according to the adaptation law~(\ref{eq:adaptive-law-network}).
Simulation results are plotted in Figure~\ref{fig:fn-adp-network}.
}{adaptation-fn-network}

\noindent {\it Additional Remarks}: \newline
\indent $\bullet$\ Leaders holding different knowledges are allowed to exist
in the same network, just like a human society may contain experts in 
different fields. As an example, consider~(\ref{eq:general-network-knowledge-leader})
again. Assume the dynamics ${\bf f}$ contains $l$ parameter sets 
${\bf a}^1, {\bf a}^2, \ldots, {\bf a}^l$ with
$$
{\bf f}({\bf x}_i,{\bf a}_i^1, \ldots, {\bf a}_i^l, t)\ =\ 
{\bf f}({\bf x}_i,{\bf a}^1, \ldots, {\bf a}^l, t)\ +\ 
\sum_{k=1}^l {\bf W}_k({\bf x}_i,t)\tilde{\bf a}_i^k
$$ 
Denoting by $\Omega^1, \Omega^2, \ldots, \Omega^l$ the followers sets corresponding to 
different knowledges, the adaptation laws are, for $k=1,2,\ldots,l$, 
$$
\dot{\bf a}_i^k\ =\ {\bf P}_i^k {\bf W}_k^T({\bf x}_i,t) 
 \sum_{j \in {\mathcal N}_i} {\bf K}_{ji}\ ({\bf x}_j - {\bf x}_i)  
\ \ \ \ \forall\ i \in \Omega^k
$$ 
States and parameters will converge under the same conditions as those
given in Theorem~\ref{th:adaptation-network}. \newline
\indent $\bullet$\ To improve the convergence rate,
the adaptation law~(\ref{eq:adaptive-law-network})
may be refined as
$$
\hat{\bf a}_i = {\bf a}_i + {\bf Q}_i {\bf W}^T({\bf x}_i,t) 
 \sum_{j \in {\mathcal N}_i} {\bf K}_{ji}\ ({\bf x}_j - {\bf x}_i)  
$$
where ${\bf Q}_i > 0$ is constant and symmetric, and ${\bf a}_i$ is defined 
by~(\ref{eq:adaptive-law-network}).
Note that in the theoretical analysis we should use a modified 
Lyapunov function
$$
V = \frac{1}{2}\ (\ {\bf x}^T {\bf L}_{\bf K} {\bf x}\ +\    
     \sum_{i \in \Omega} \tilde{\bf a}_i^T {\bf P}^{-1}_i \tilde{\bf a}_i\ )
    \ +\ \sum_{i \in \Omega} \int_0^t {\bf z}_i^T {\bf Q}_i {\bf z}_i dt
$$  
where $\tilde{\bf a}_i = {\bf a}_i - {\bf a}$ and
$\displaystyle {\bf z}_i = {\bf W}^T({\bf x}_i,t) 
 \sum_{j \in {\mathcal N}_i} {\bf K}_{ji}\ ({\bf x}_j - {\bf x}_i)$. \newline
\indent $\bullet$\ The number of leaders in a knowledge-based network can be
arbitrary. At the limit all elements could be adaptive, i.e., there is no leader at all,
in which case they may converge to any odd parameter set depending on initial conditions.
While all states will still converge together, the desired individual behaviors (such as
oscillations) may not be preserved. \newline
\indent $\bullet$\ Synchronization conditions derived in 
Theorem~\ref{th:adaptation-network} are
very similar to those in~\cite{wei03-2, wei03-1} for coupled 
networks without any leader or adaptation.
Note that if the condition~(\ref{eq:negative-semi-definite}) is true, 
$\forall$ neighbored $i, j$, ${\bf x}_i - {\bf x}_j$ 
are bounded. Thus the boundedness of the states are simply
determined by the Input-to-State Stability~\cite{khalil} of the system
$
\dot{\bf y} =  {\bf f}({\bf y},{\bf a}, t) + {\bf u}
$
where the input ${\bf u}$ is bounded. \newline 
\indent $\bullet$\ The condition~(\ref{eq:parameter-converge-condition}) is 
true if the stable system behaviors are sufficiently rich or persistently 
exciting. This is the 
case, for instance, when the individual elements are oscillators, where
the possibilities that any component of ${\bf x}_i$ converges to zero can 
be excluded by dynamic analysis showing that zero is an unstable state. \newline
\indent $\bullet$\ Both power leaders and knowledge leaders could be virtual, which
is common in animal aggregate motions. For instance, a landmark may be
used as a virtual power leader. Similarly, when hunting, an escaping
prey could specify both the where and the how of the movement.

%
\section{Pacific Coexistence} \label{sec:mixture-leaders}
Different types of leaders can co-exist in the same network.
A power leader could be also a knowledge leader, or conversely,
as we illustrated in Figure~\ref{fig:power-leader}(c),
a power leader guiding the
direction may use state measurements from its neighbors to adapt its
parameters to the values of the knowledge leaders.

Consider the power-based leader-followers network~(\ref{eq:leader-followers})
again, assuming the dynamics ${\bf f}$ contains a parameter set ${\bf a}$.
There are knowledge leaders holding the fixed value ${\bf a}$ and knowledge 
followers using adaptation to learn. If $0 \in \Omega$, the set of the knowledge
followers, we have
$$
\dot{\bf a}_0\ =\ {\bf P}_0 {\bf W}^T({\bf x}_0,t) 
 \sum_{i=1}^n \gamma_i\ {\bf K}_{0i}\ ({\bf x}_i - {\bf x}_0)  
$$
while if $i \in \Omega$ with $i=1,\ldots,n$, 
$$
\dot{\bf a}_i\ =\ {\bf P}_i {\bf W}^T({\bf x}_i,t) 
 ( \sum_{j \in {\mathcal N}_i} {\bf K}_{ji}\ ({\bf x}_j - {\bf x}_i) +
  \gamma_i\ {\bf K}_{0i}\ ({\bf x}_0 - {\bf x}_i) \ ) 
$$

To prove state convergence, first we define several Laplacian matrices for a 
power-based network structure: \newline
$\bullet$\ ${\bf L}_{\bf K}$, the weighted Laplacian of the followers network.\newline
$\bullet$\ $\vec{\bf L}_{\bf K}$, the weighted Laplacian of the whole network, which 
is non-symmetric since we have uni-directional links between the leader and the 
followers. Thus,
$$
\vec{\bf L}_{\bf K} = \left[ \begin{array}{cc} {\bf 0} & {\bf 0} \\                         
                   -{\bf b}  & {\bf C} \end{array} \right]
\ \ \ \ \mathrm{where} \ \ \
{\bf b} = \left[ \begin{array}{c} \vdots \\ \gamma_i\ {\bf K}_{0i} \\ \vdots  \end{array} \right] ,
\ \ 
{\bf C} = {\bf L}_{\bf K} +  {\bf I}^n_{\gamma_i {\bf K}_{0i}}
$$
${\bf C}$ is positive definite if the whole network
is connected. \newline
$\bullet$\ $\bar{\bf L}_{\bf K}$,  the weighted Laplacian of the 
whole network which we consider as an undirected graph. Thus,
$$
\bar{\bf L}_{\bf K} = {\vec{\bf L}_{\bf K}}^T+
\left[ \begin{array}{cc} \displaystyle \sum_{i=1}^n \gamma_i {\bf K}_{0i} & {\bf 0} \\                         
                    -{\bf b}  & {\bf 0} \end{array} \right]
$$

Define the Lyapunov function
$$
V = \frac{1}{2}\ (\ {\bf x}^T \bar{\bf L}_{\bf K} {\bf x}\ +\ 
         \sum_{i \in \Omega} \tilde{\bf a}_i^T {\bf P}^{-1}_i \tilde{\bf a}_i\ )
$$
We can show that
$$
\dot{V} 
\ =\ {\bf x}^T \bar{\bf L}_{\bf K}\ (\  
      \left[ \begin{array}{c} {\bf f}({\bf x}_1,{\bf a},t) \\ \ldots \\  
                {\bf f}({\bf x}_n,{\bf a},t) \end{array} \right]
        - \vec{\bf L}_{\bf K}{\bf x}\ ) 
\ =\ {\bf x}^T\ (\ \bar{\bf L}_{\bf K \Lambda} - 
        {\vec{\bf L}_{\bf K}}^T\ \vec{\bf L}_{\bf K} \ )\ {\bf x}
$$
where $\bar{\bf L}_{\bf K \Lambda}$ is defined similar as~(\ref{eq:l-k-lambda}),
except that here the incidence matrix is based on the whole network.
See Appendix~\ref{ap:leader-combination-proof} for the condition for
$\bar{\bf L}_{\bf K \Lambda} - {\vec{\bf L}_{\bf K}}^T\ \vec{\bf L}_{\bf K}$ to 
be negative semi-definite. Following the same proofs as those in 
Sections~\ref{sec:knowledge-leader}, this then implies that all the states 
${\bf x}_i$, $i=0,1,\ldots,n$ will converge together asymptotically. Parameter 
convergence conditions are also the same.

%
\section{Concluding Remarks} \label{sec:conclusion}
We studied two types of leaders for networked systems, which spread
desired orientation or dynamics through distributed interactions. Synchronization
conditions were derived, as well as the parameter convergence conditions for
knowledge-based leader-followers networks.


\section{Appendices}
For notational simplicity, we show the derivations for the case $m=1$.

\subsection{Proof of Lemma ${\bf 1}$} 
\label{ap:negative-semi-definite-proof}
Notice that $0$ is always one of the eigenvalues of
${\bf L}_{\bf K \Lambda} - {\bf L}_{\bf K}^2$, with one corresponding eigenvector
${\bf v} = [1, 1, \ldots, 1]^T$. According to Weyl's Theorem~\cite{horn},
$$
\lambda_{n-k+1}({\bf L}_{\bf K \Lambda} - {\bf L}_{\bf K}^2)\ \le\
\lambda_n({\bf L}_{\bf K \Lambda})-\lambda_k({\bf L}_{\bf K}^2)
$$
where $k=1,2,\ldots,n$, and the eigenvalues $\lambda_i$ are arranged in increasing 
order for $i=1,2,\ldots,n$. This implies that, $\forall k>1$,
$\lambda_{n-k+1}({\bf L}_{\bf K \Lambda} - {\bf L}_{\bf K}^2)<0$
if
\begin{equation} \label{eq:condition-negative-semi}
\lambda_n({\bf L}_{\bf K \Lambda})\ <\ \lambda_2({\bf L}_{\bf K}^2)
\end{equation}
Therefore, 
$
\lambda_n({\bf L}_{\bf K \Lambda} - {\bf L}_{\bf K}^2)=0
$, i.e., ${\bf L}_{\bf K \Lambda} - {\bf L}_{\bf K}^2$ is negative semi-definite.

Denote
$ \displaystyle
\max_k \lambda_{max}({\bf K \Lambda})_{ijs} = \bar{\lambda}
$.
If $\bar{\lambda} \le 0$, we have 
$\lambda_n({\bf L}_{\bf K \Lambda}) \le 0$ and both the 
conditions~(\ref{eq:condition-negative-semi}) 
and~(\ref{eq:negative-semi-definite}) are always true; if
$\bar{\lambda}>0$,
$$
\lambda_n({\bf L}_{\bf K \Lambda})\ \le\  \bar{\lambda}\ \lambda_n({\bf L})
$$
where ${\bf L}$ is the graph Laplacian matrix. Considering the fact that
$
\lambda_2({\bf L}_{\bf K}^2) = \lambda_2^2({\bf L}_{\bf K})
$,
condition~(\ref{eq:negative-semi-definite}) is sufficient to guarantee
(\ref{eq:condition-negative-semi}).

For a real symmetric matrix, the state space has an orthogonal basis consisting of all
eigenvectors. Without loss generality, we assume there is such an orthogonal eigenvector 
set, $\{{\bf v}_1, {\bf v}_2, \ldots, {\bf v}_n\}$,
of ${\bf L}_{\bf K \Lambda} - {\bf L}_{\bf K}^2\ $, 
where ${\bf v}_n = [1, 1, \ldots, 1]^T$
is the only zero eigenvector. For any ${\bf x}$, we have
$$
{\bf x} = \sum_{i=1}^n k_i{\bf v}_i\ \ \  \mathrm{and}\ \ \ 
{\bf x}^T\ (\ {\bf L}_{\bf K \Lambda}-{\bf L}_{\bf K}^2\ )\ {\bf x}
= \sum_{i=1}^{n-1} \lambda_i k_i^2 {\bf v}_i^T {\bf v}_i
$$
Since the eigenvalue $\lambda_i<0$ $\forall i<n$,
${\bf x}^T\ (\ {\bf L}_{\bf K \Lambda}-{\bf L}_{\bf K}^2\ )\ {\bf x}=0$ if 
and only if ${\bf x} =  k_n{\bf v}_n$, that is,
${\bf x}_1 = {\bf x}_2 = \cdots = {\bf x}_n$.

In case $m>1$, we can follow the same proof except that zero eigenvalue here has $m$ 
multiplicity, and the corresponding eigenvectors 
$\{{\bf v}_1, {\bf v}_2, \ldots, {\bf v}_m\}$ are linear combinations of the orthogonal set 
$[{\bf I}, {\bf I}, \ldots, {\bf I}]^T$ where 
${\bf I} \in \mathbb{R}^{m \times m}$ is identity matrix.

%
\subsection{Network with Both Leaders} 
\label{ap:leader-combination-proof}
Similarly to the proof in~\ref{ap:negative-semi-definite-proof},
$\bar{\bf L}_{\bf K \Lambda} - {\vec{\bf L}_{\bf K}}^T\ \vec{\bf L}_{\bf K}$
is negative semi-definite if
$$
\lambda_{n+1}(\bar{\bf L}_{\bf K \Lambda})\ <\ 
 \lambda_2( {\vec{\bf L}_{\bf K}}^T\ \vec{\bf L}_{\bf K} )
$$ 
and its only eigendirection for the zero eigenvalue is thus
${\bf v} = [1, 1, \ldots, 1]^T$. Since
$$
{\vec{\bf L}_{\bf K}}^T\ \vec{\bf L}_{\bf K} = 
\left[ \begin{array}{cc} {\bf b}^T {\bf b} & -{\bf b}^T {\bf C} \\                         
                   -{\bf C}{\bf b}  & {\bf C}^2 \end{array} \right]
$$
we have 
$$
\lambda_2( {\vec{\bf L}_{\bf K}}^T\ \vec{\bf L}_{\bf K} )
\ge \lambda_1( {\bf C}^2 ) = \lambda_1^2( {\bf C} )
$$
according to the Interlacing Eigenvalues Theorem for bordered matrices~\cite{horn}.
Thus a sufficient condition to guarantee negative semi-definite is 
\begin{equation} \label{eq:condition-negative-semi-combination}
 \lambda_1^2( {\bf C} ) > \lambda_{n+1}(\bar{\bf L}_{\bf K \Lambda})
\end{equation}
This condition is similar to the one we derived in Theorem~\ref{th:power-leader} 
for synchronization of pure power-based leader-followers network.
Assuming all the coupling strengths are identical with value $\kappa$,
condition~(\ref{eq:condition-negative-semi-combination}) becomes
$$
\kappa\ >\ \frac{\lambda_{n+1}(\bar{\bf L})}
                {\lambda_1^2({\bf L} + {\bf I}^n_{\gamma_i})}\  
\max \frac{\partial {\bf f}}{\partial {\bf x}} 
                        ({\bf x}_i, {\bf a}, t)
$$

%
%
\renewcommand{\baselinestretch}{1.}

\end{document}